**Cleaning graphene : a first quantum/classical molecular dynamics approach**

**L. Delfour[1], A. Davydova[2], E. Despiau-Pujo[2], G. Cunge[2], D.B. Graves[3], L. Magaud[1]**

[1] Institut Néel, CNRS/UJF-Grenoble 1, 25 avenue des Martyrs, 38054 Grenoble, France

[2] LTM, CNRS/UJF-Grenoble1/CEA, 17 avenue des Martyrs, 38054 Grenoble, France

[3] University of California at Berkeley, Department of Chemical & Biomolecular Engineering, Berkeley, CA 94720, USA

**E-mail:** laurence.magaud@grenoble.cnrs.fr **;** emilie.despiau-pujo@cea.fr

**abstract:**

Graphene outstanding properties created a huge interest in the condensed matter community and unprecedented fundings at the international scale in the hope of application developments. Recently, there have been several reports of incomplete removal of the polymer resists used to transfer as-grown graphene from one substrate to another, resulting in altered graphene transport properties. Finding a large-scale solution to clean graphene from adsorbed residues is highly desirable and one promising possibility would be to use hydrogen plasmas. In this spirit, we couple here quantum and classical molecular dynamics simulations to explore the kinetic energy ranges required by atomic hydrogen to selectively etch a simple residue – a $CH_3$ group - without irreversibly damaging the graphene. For incident energies in the 2-15 eV range, the $CH_3$ radical can be etched by forming a volatile $CH_4$ compound which leaves the surface, either in the $CH_4$ form or breaking into $CH_3+H$ fragments, without further defect formation. At this energy, adsorption of H atoms on graphene is possible and further annealing will be required to recover pristine graphene.

## 1. Introduction

Graphene exceptional properties open exciting opportunities in many fields, such as energy storage, electronics, photonics, coating [1]. While some of these applications can work with a defected material, many other require a nearly perfect graphene. During the last years, an intense research activity has focused on how defects can affect or alter graphene properties [2,3] and graphene cleaning is becoming a crucial issue, a mandatory step for further application development.

Defects appear during the different steps required to grow and transfer graphene or create devices. Indeed, one way to grow and obtain large-scale high quality graphene is to use metallic surfaces [4]. Then, for many applications it has to be transferred to another substrate. This is commonly performed using PMMA resist – Poly(methyl methacrylate) – in order to ensure rigidity while removing the metallic substrate. After transfer, the polymer is removed by thermal annealing and the use of solvents, but full removal cannot be achieved. PMMA thermal decomposition leads to the formation of residues adsorbed on graphene and modify its properties, inducing p-doping and degrading its electronic mobility [5-7].

Hydrogen plasma provides one way to pattern graphene and it has been shown that, depending on the temperature, it can selectively etch ribbon edges or create holes in the basal plane [8,9]. Recently, the use of hydrogen plasmas has been proposed [10] to clean graphene from PMMA resist residues. In low-temperature plasma processing, the surfaces are exposed simultaneously to an isotropic flux of thermal radicals (e.g. H atoms in a $H_2$ plasma) and to an anisotropic flux of energetic ions (e.g. H+), which can result in the selective etching of a material. To clean graphene surfaces in $H_2$ plasmas, the range of kinetic energies needed by the plasma species to selectively etch the residues must be determined. One important question one has to answer is whether the impinging hydrogen species (especially ions) would create new defects in graphene and whether they can be cured or not. This is a completely new and tough issue to model from scratch. To start tackling it, we use a simple description of a PMMA residue, a methyl group, and calculate the effect of an impinging H atom arriving on the C atom of the $CH_3$ group at normal incidence with respect to the graphene plane, as a function of its incident energy and for various surface temperatures. Modeling plasma ions impacts using fast neutral atoms at normal incidence is a reasonable approximation because in low-pressure plasma reactors, ions are accelerated through the plasma sheath perpendicularly to the surfaces, so that H+ ions reach the target sample with a normal incidence [11]. Furthermore, when a plasma ion reaches a conductive surface, it is generally assumed to be neutralized by an auger mechanism [12].

Beyond the fact that $CH_3$ radicals can result from PMMA thermal decomposition, etching of methyl groups adsorbed on graphene is also of strong interest since high quality, single layer graphene growth can be obtained through successive cycles of growth using an organic compound gas and etching by hydrogen [13].

Following what we did for elementary hydrogen interactions with surface-clean graphene [14], we couple here quantum (QMD) and classical (CMD) molecular dynamics simulations to study the effect of energetic H atoms impacting a $CH_3$ residue adsorbed on graphene. While quantum molecular dynamics treats in an accurate way the electronic part of the problem, it remains a heavy computational approach and thus the number and duration of calculated trajectories is limited. On the other hand, classical molecular dynamics uses an empirical potential form to model the interactions between atoms. While less accurate, it enables the sampling of a much larger number of trajectories. Coupling the two approaches allows to check further the accuracy of the empirical potential against ab initio results and validate it in a more complex case than that of Ref 14. One main difficulty here comes from the fact that the potential has to describe $sp^2$ as well as $sp^3$ bonded carbon atoms together with the C-H interaction. Once validated, this empirical potential is used to address more complex cases including longer dynamics, fine sampling of the trajectories as a function of the incident H energy, effect of surface temperature.

At 0 K, identical mechanisms are predicted by both QMD and CMD for increasing incident energy:  reflection and chemical etching. However, plasma processing usually takes place at temperatures closer to 300K, or even higher, and thermal vibrations of the substrate atoms may modify the reaction probabilities for the different processes. Thus, CMD is then used to evaluate the influence of graphene temperature on the interaction mechanisms.

All simulations show that at moderate incident energy, the radical is etched. This happens through the formation of a volatile $CH_4$ compound that leaves graphene without new defect creation. At this energy, adsorption of H atoms on graphene is possible as shown in Ref. 14 and further annealing will be required to recover pristine graphene.

The paper is organized as follows: computational details are given in part 2, the different reaction mechanisms at 0K are discussed in part 3 and surface temperature effect is presented in part 4. Conclusions are discussed in part 5.

## 2. Computational Approaches

### 2.1. Quantum Molecular Dynamics (QMD)

Quantum Molecular Dynamics calculations are performed using the code VASP [15,16]. A $CH_3$ radical adsorbed at the center of the graphene cell has first been fully relaxed (Figure 1). Atomic hydrogen is then sent at normal incidence on the C atom of the $CH_3$ radical for a chosen set of initial energies. All calculations are performed in the NVE statistical ensemble at 0K, with a time step equal to 0.1 fs. Newton's equations of motion are integrated using a Verlet algorithm [17].

We are using the Perdew-Wang generalized gradient functional (PW91) [18] and the PAW approach [19]. A 6x6x1 k-points grid is used and the supercell cell has a 24.7x21.4x18 $Å^3$ dimension. It contains 200 +1 C atoms and 4 H atoms. The size along the z direction is large enough so that the image layers would not interact.

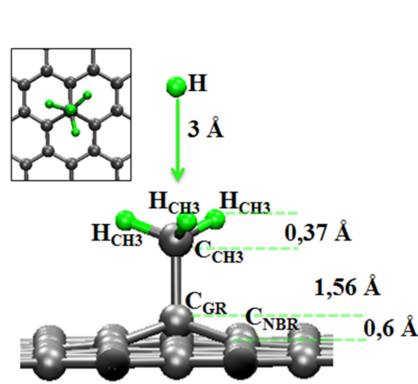

Figure 1: Initial configuration used in QMD and CMD calculations at 0K (side-view). Insert: top-wiew.

### 2.2. Classical Molecular Dynamics (CMD)

Classical molecular dynamics (CMD) refers to a class of simulation that solves Newton's equations of motion for a system of N interacting particles, treating each atoms as a classical point and modeling the quantum effects from electrons by an interatomic potential energy function. In this work, we use the 2nd generation C-H REBO potential developed by Brenner and coworkers [20]. This potential has been widely used to model solid carbon and various hydrocarbon systems.  A description of our calculation methods can be found in a former paper [14], where we tested the REBO potential against DFT calculations to study elementary processes of graphene-hydrogen interaction.

CMD simulations are performed in the microcanonical NVE ensemble. The Velocity-Verlet algorithm is used to calculate positions and velocities, with a timestep equal to 0.1 fs. The graphene surface is represented as a single-layer sheet with periodic boundary conditions along the Ox and Oy axes to mimic a semi-infinite plane. Four atoms are fixed on the edges to anchor the simulation cell. The initial configuration contains 260 atoms with a surface area of 670 Å$^2$ (Lx =19.67 Å, Ly = 34.08 Å). In order to model a CH$_3$ radical absorbed on the graphene surface, a methyl group is created in the middle of the cell and allowed to relax for 100 ps to get its equilibrated structure. Atomic hydrogen with energy E$_i$ between 0.1 eV and 25 eV is sent at normal incidence, from 3 Å above the full configuration, as presented in Figure 1. After injecting the H atom above the CH$_3$ radical, the trajectories of all atoms are followed during ~ 1 ps. At the end of each impact, the surface reaction is observed and analyzed. The original temperature of the whole system (Tcell) is set to 0K ; the simulation cell is then quenched at 300K and 600K to study the influence of the surface temperature.

## 3. Reaction mechanisms at 0K

In the following, the impact of an incident H atom with varying initial velocities is described. Various mechanisms (H reflection, CH$_4$ formation and desorption, CH$_3$ etching or sputtering) are observed when the incident kinetic energy is increased. At 0K and for small and intermediate incident energy, DFT and CMD calculations predict similar mechanisms but for shifted energy ranges.

### *3.1. Initial configuration*

The geometry at 0K of a methyl group adsorbed on graphene is shown in Figure 1, together with the notations used throughout the text. DFT and CMD fully relaxed geometries are in good agreement. The $CH_3$ group is rotated with respect to the graphene. The length of the 3 $C_{CH3}$-$H_{CH3}$ bonds are identical and equal to 1.1 Å. The radical carbon atom, $C_{CH3}$, is located 1.56 Å above the carbon atom belonging to the graphene surface ($C_{GR}$). The graphene carbon atom, $C_{GR,}$ raises 0.6 Å above the graphene basal plane, stretching the bonds with its nearest carbon neighbors $C_{GR}$-$C_{NBR}$ up to 1.51 Å. The $CH_3$ radical chemisorption depletes the local π bonds in the graphene structure, thus $C_{GR}$-$C_{NBR}$ bonds become weaker. It leads to a partial transition of the local bonding from $sp^2$ to $sp^3$ in agreement with other calculations [21-23].

 A hydrogen atom with different incident energies is inserted into the system 3 Å above the $CH_3$ radical. This distance is large enough so that in the starting configuration, no interaction exist between the incident H and the $CH_3$ group.

### 3.1. Reflection

Results from CMD simulations show that for $E_i$ < 3.7 eV, the impacting H atoms are systematically reflected from the surface without modifying the structure of the $CH_3$ radical bound to the graphene. In order to illustrate such reaction, we consider the impact of a 1 eV H atom onto the $CH_3$ radical. Figure 2 shows the evolution of interatomic distances as a function of time before and after the impact. The solid red curve corresponds to the distance between the impinging H atom and the $C_{CH3}$ atom, the dashed blue and dotted red ones to the $C_{CH3}$-$C_{GR}$ and $C_{CH3}$-$H_{CH3}$ bonds, respectively. One can see that during the impact, the H atom comes close (~ 1.6 Å) to $C_{CH3}$ (the carbon atom belonging to the $CH_3$ radical)  and then gets reflected. The target radical reacts to this low energy impact with small fluctuations of its $C_{CH3}$-$H_{CH3}$ bond lengths around their average value of 1.1 Å. At the same time, the $C_{CH3}$ atom is slightly pushed down after the H impact; thus the $C_{CH3}$-$C_{GR}$ bond is shortened and then shows small variations around its initial length (1.56 Å). Contrary to CMD results, DFT calculations at 1eV do not show H reflection from the surface but another mechanism - the formation and desorption of a $CH_4$ molecule. This mechanism is also obtained with CMD but for slightly higher energies [3.7-8 eV], as discussed in the next paragraph. Based on this reflection result obtained with CMD, we performed QMD calculations for decreasing incident energies and indeed found the occurrence of a reflection but at much smaller incident energy (0.07 eV). It is a first evidence of the technical interest of coupling classical and quantum molecular dynamics: using the classical approach to cover a broader range of configurations and checking interesting ones with DFT.

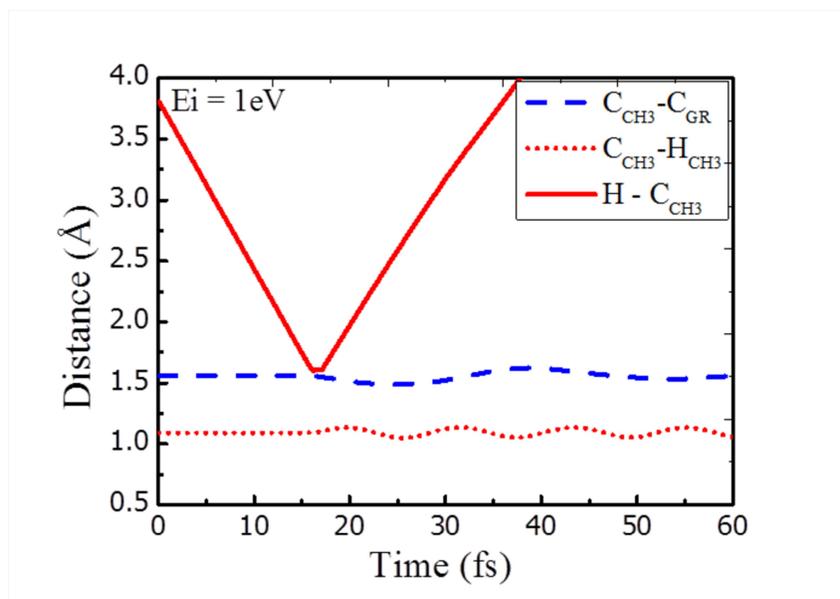

Figure 2:

CMD calculation of bond length variations as a function of time, when a H atom impacts onto the CH₃ radical with an incident energy of 1eV. Red solid, blue dashed and red dotted curves correspond to H-$C_{CH3}$, $C_{CH3}$-$C_{GR}$ and $C_{CH3}$-$H_{CH3}$ distances respectively.

### *3.2. CH₄ formation and desorption*

QMD calculations predict that a hydrogen atom with an initial energy equal to 1 eV is adsorbed by the methyl group to form a volatile CH₄ molecule, which then desorbs from the graphene surface. Figure 3 presents the evolution of the CH₃ target bond lengths before and after the H impact. One can see that when the H atom comes close to the radical $C_{CH3}$ atom, a bond is formed and oscillates around ~ 1.1 Å, which is the equilibrium value for C-H bonds in methane molecules. Once the CH₄ molecule is formed (from 10 fs), the distance between $C_{GR}$ and $C_{CH3}$ continuously increases, which indicates that the CH₄ molecule leaves the graphene surface. At the end of the simulation time, the volatile molecule is already 3.5 Å away from the graphene surface and has thus desorbed.

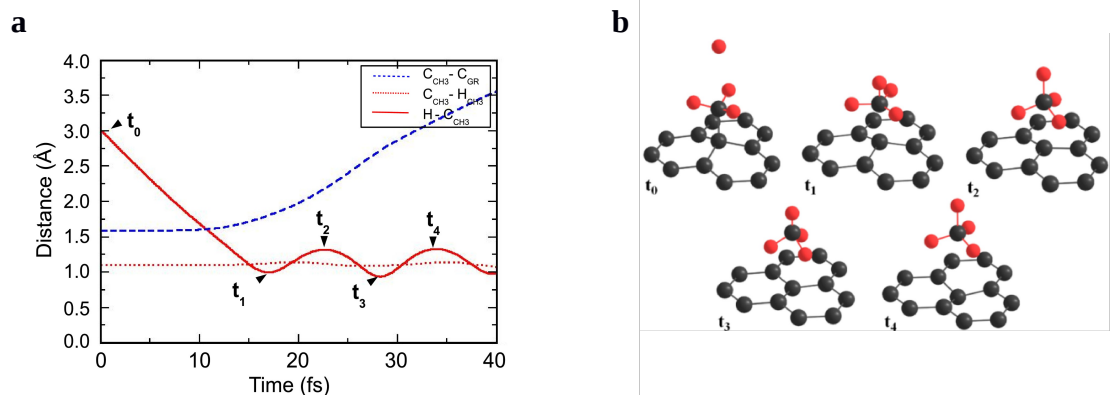

Figure 3 : (a) QMD calculation of bond lengths describing H-$C_{CH3}$ (red), $C_{CH3}$-$C_{GR}$ (dashed blue), $C_{CH3}$-$H_{CH3}$ (red dots) as a function of time, for a 1 eV incident energy. (b) Snapshots of the QMD simulation for specific times of 0 fs, 17 fs, 22.7 fs, 28.3 fs and 33.9 fs. Shown by arrows in (a).

The $C_{CH3}$-$H_{CH3}$ bond length slightly oscillates after H absorption with an amplitude that is much smaller than the $C_{CH3}$-H bond length variations. Figure 3b shows the main phases of CH$_4$ formation and desorption through snapshots of the simulation, from times t0 to t4 chosen to highlight the H-C$_{CH3}$ behavior. Interaction is no more negligible for a H-C$_{CH3}$ distance equal to 1.9 Å: As the incident H atom approaches, the carbon atom of the group (C$_{CH3}$) raises toward the projected atomic hydrogen, while the three H$_{CH3}$ atoms which compose the radical move downwards toward the surface by 0.55 Å (t1, t2, t3), in order to accommodate the additional hydrogen atom. The formation of the CH$_4$ molecule strongly weakens the interaction with graphene. The movement of the three H$_{CH3}$ can be compared to an "umbrella inversion" and bring them close to the graphene surface. The resulting repulsion leads to the desorption of the newly formed CH$_4$ molecule which is not bound to graphene by a C-C bond anymore. At the same time, the carbon atom C$_{GR}$ belonging to the graphene layer sinks by 0.3 Å below the surface and starts to oscillate.

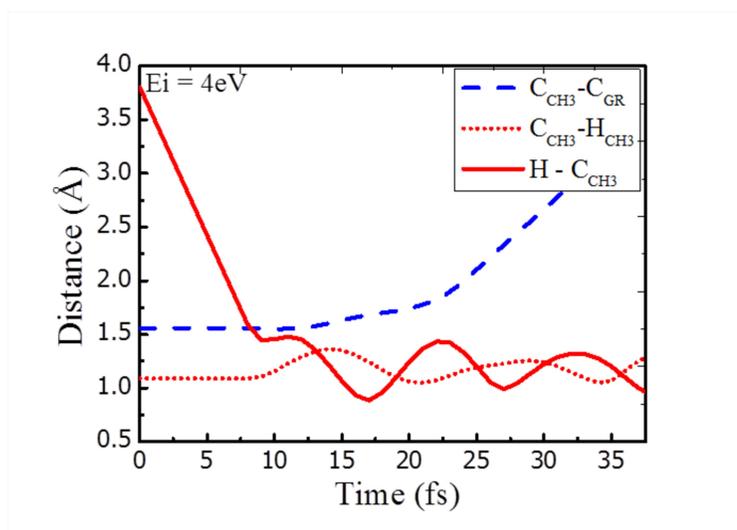

Figure 4: CMD calculation of bond lengths describing H-C$_{CH3}$ (red), C$_{CH3}$-C$_{GR}$ (dashed

blue), $C_{CH3}$-$H_{CH3}$ (red dots) as a function of time, for a 4 eV incident energy.

CMD simulations predict exactly the same surface reaction but for a [3.7-8 eV] energy range, which is higher than the DFT results. Figure 4 shows the time variations of bond distances obtained from CMD when a H atom impinges onto $C_{CH3}$ with 4eV. As previously observed in Figure 3, the H atom binds to the $CH_3$ radical. The H-$C_{CH3}$ bond length first experiences strong fluctuations and then reaches equilibrium, oscillating around its average value of 1.1 Å. Then the $C_{CH3}$-$H_{CH3}$ bonds respond to the H absorption and also start to fluctuate with a constant amplitude of ~ 0.35 Å. These fluctuations initiate the desorption process of the formed methane molecule. In Figure 4, the constant increase of the $C_{CH3}$-$C_{GR}$ distance demonstrates that the volatile product leaves the graphene surface. After 40 fs, the $CH_4$ molecule is already 4 Å above the graphene surface.

QMD and CMD show a similar formation/desorption mechanism for the $CH_4$ molecule. Both approaches also indicate that at the end of the simulation, neither vacancy nor defects are observed on the graphene layer, which retrieves its original geometry.

QMD calculations for an incident energy close to 2.5 eV show a very similar behavior. The amplitude of the $C_{CH3}$-H distance oscillations is just much larger (1.25 Å instead of 0.5 Å). This 2.5 eV case is close to the limit where the $CH_4$ molecule first forms and then breaks into a $CH_3$ radical and an H atom as it is shown in the next section.

### *3.3. CH$_3$ formation*

For higher incident energies, another reaction mechanism leading to the methyl removal from graphene is observed with both classical and quantum approaches, but again for different energy ranges (QMD at 4 eV, CMD in the [9-11.6] eV energy range).  Figure 5 shows the evolution of the distances calculated by QMD for this velocity and the simulation snapshots which can be directly compared to the ones in Figure 3. The projected atomic H comes 0.18 Å closer to $C_{CH3}$ than for an incident energy equal to 1 eV. The bond is then strongly compressed and the H bounces back with an energy sufficient to break the $C_{CH3}$-H bond. As observed previously, the etching of the $CH_3$ group proceeds through an "umbrella inversion" of the three $H_{CH3}$ atoms. It proceeds more slowly than previously observed for $CH_4$, since at 37 fs $C_{CH3}$ is only 2.42 Å away from $C_{GR}$ (3.36 Å in the previous case). The $C_{CH3}$-$H_{CH3}$ bonds also show slightly larger amplitude oscillations of about 0.04 Å (instead of 0.03 Å).

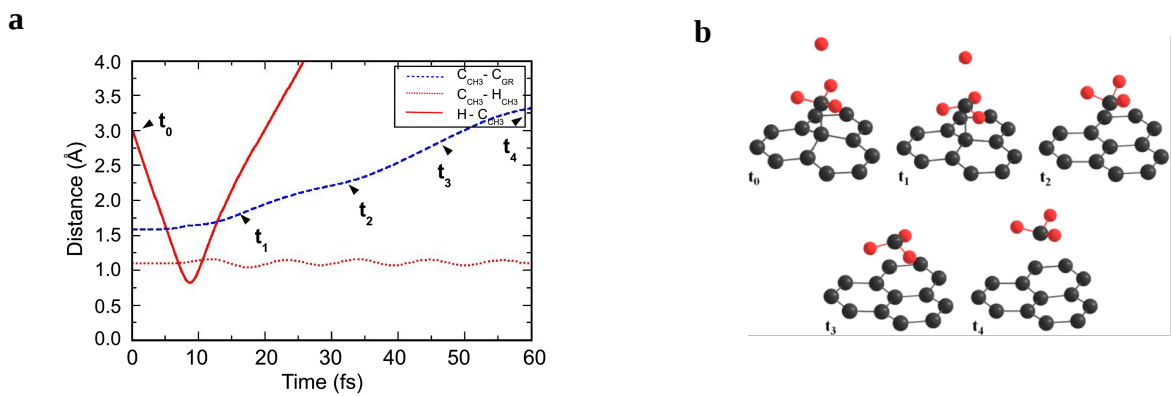

Figure 5 :  (a) QMD calculation of relevant bond lengths describing H-$C_{CH3}$ (red full line),

$C_{CH3}$-$C_{GR}$ (blue dashed line), $C_{CH3}$-$H_{CH3}$ (red dots) as a function of time, for a 4 eV incident energy. (b) Snapshots of the QMD simulation for specific times of 0 fs, 17 fs, 33 fs, 47 fs and 68.3 fs.

The set of curves shown in Figure 6 presents the behavior of bond distances calculated from CMD after H has impacted with $E_i = 9$ eV. Here the impacting H atom comes close to the $CH_3$ radical and then binds to it and forms a $CH_4$ molecule. The excess kinetic energy from the impacting H is transferred to the neighbor atoms in the $CH_3$ group and to the graphene atoms. Thus both $C_{CH3}$-$H_{CH3}$ and $C_{CH3}$-$C_{GR}$ bonds experience stress and fluctuations initiating the desorption process of the newly formed $CH_4$. As a consequence, after 10 fs, the $C_{CH3}$-$C_{GR}$ bond stretches and breaks, allowing the volatile methane product to leave the surface. However, once $CH_4$ has left the graphene surface, it immediately disintegrates into a single H and a $CH_3$ group. This molecular dissociation is possible due to the excess kinetic energy carried by H, which induces the desorption of a $CH_4$ molecule in a highly excited vibrational state (pre-dissociated). In both classical and quantum cases, the etch product is $CH_3$ and the graphene surface is free of any defects after the methyl removal by the atomic H.

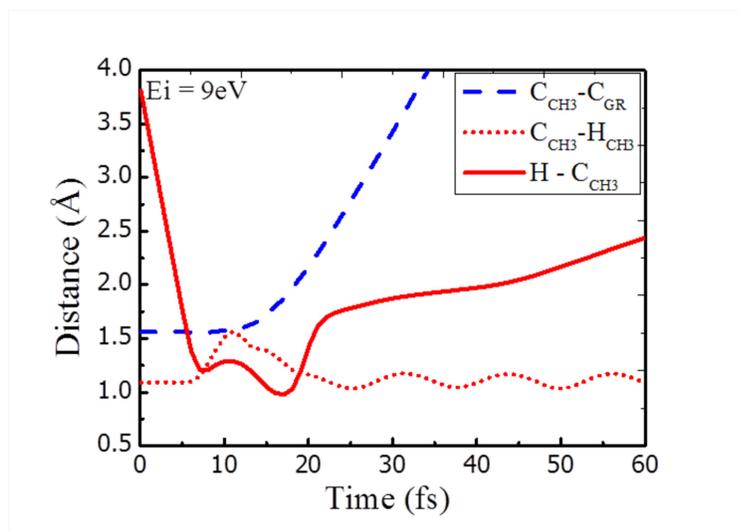

Figure 6: CMD calculation of bond lengths as a function of time when a H atom impacts onto the CH₃ radical with an incident energy equal to 9 eV. Colors are identical to the ones in Figure 5.

### 3.4. Reflection or sputtering events at high energies

Eventually, if the energy of the incident H atom is increased further, other reaction mechanisms are observed in both QMD and CMD calculations. In this high incident energy case only, the two mechanisms differ somehow: QMD calculations show a reflection of the H atom with a methyl group that remains adsorbed on graphene - calculations performed for initial energies equal to 8, 20 and 30 eV-, while CMD calculations show a reflection of the H atom and an eventual sputtering of the $H_{CH3}$ atoms with the single $C_{CH3}$ atom remaining adsorbed on graphene.

**b**

**a**

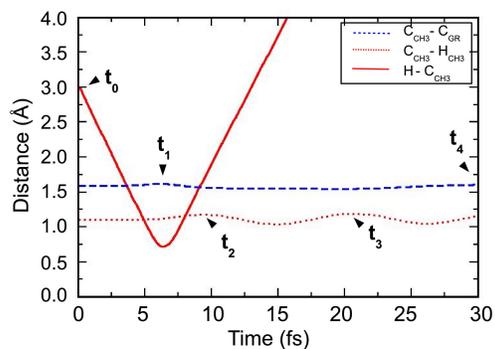

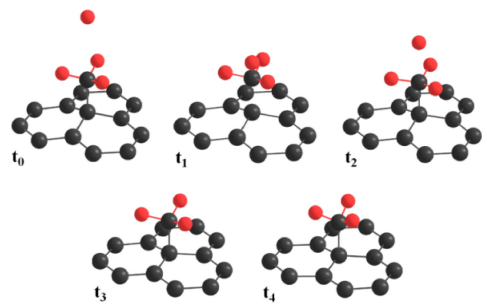

Figure 7: (a) QMD calculation of bond lengths describing H-$C_{CH3}$ (red), $C_{CH3}$-$C_{GR}$ (blue), $C_{CH3}$-$H_{CH3}$ (red dots) as a function of time, for a 8 eV incident energy. (b) Snapshots of the QMD simulation for specific times of 0 fs, 6.2 fs, 10 fs, 20 fs and 27.9 fs.

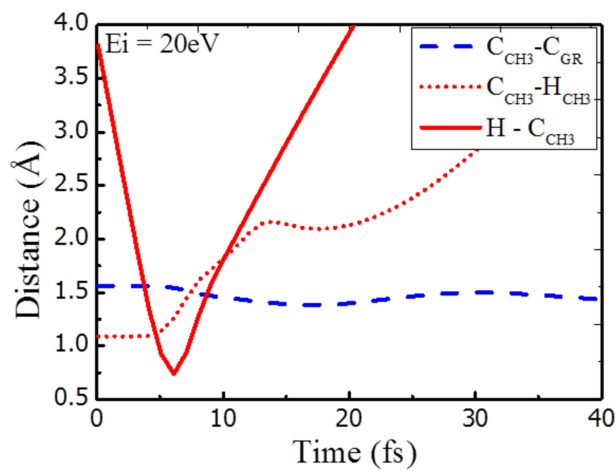

Figure 8: CMD calculation of bond distances variation as a function of time when an H atom impacts onto the $CH_3$ radical with an incident energy equal to 20 eV. Colors are identical to the ones in Figure 7.

In QMD calculation, the amplitude of the bond lengths variations for an initial energy of 8 eV are larger within the $CH_3$ radical (0.134 Å) than between the radical and the surface ($C_{CH3}$-$C_{GR}$) (0.069 Å at 18.2 fs) ; however, it does not lead to the sputtering of the $H_{CH3}$ atoms as predicted by CMD .

Indeed for high incident H energies (> 11.6eV), CMD calculations predicts not only the reflection of the incident H (like in QMD) but also another possible reaction path: the sputtering of one or more $H_{CH3}$ atoms from the $CH_3$ radical. For example, Figure 8 shows the impact of a 20 eV H atom onto the methyl radical as calculated by CMD. In this case, due to its high kinetic energy, the impacting hydrogen is reflected very rapidly but transfers a significant part of its kinetic energy to the target $CH_3$ radical (the velocity of the reflected H atom changes after 8 fs as shown in Figure 8). As a consequence, the three $H_{CH3}$ atoms are sputtered away from $C_{CH3}$ as three single H atoms, leaving the single $C_{CH3}$ atom chemisorbed on the graphene basal plane. This sputtering mechanism is coherent with the QMD calculations, which indicate that most of the energy of the impact goes into vibrations of the $C_{CH3}$-$H_{CH3}$ bonds; however, in QMD calculations, the amount of energy transferred from the H atom to the methyl group is not large enough to sputter the $H_{CH3}$ atoms. In addition, the incident H impact also initiates small fluctuations of the $C_{GR}$-$C_{CH3}$ bond with an amplitude of 0.11 Å.

Although the mechanisms differ slightly depending on the energy, both classical and quantum results show no etching of the whole $CH_3$ group. At 0K and for incident kinetic energies higher than 8eV in QMD (12eV in CMD), the $C_{CH3}$ atom of the residue always remains bound to the graphene basal plane. This suggests that too high incident energies may not be suitable to clean graphene from simple residues.

To conclude on these calculations performed at 0K, similar $H/CH_3$ interaction mechanisms are predicted by QMD and CMD for increasing incident energy: H reflection, $CH_4$ formation and desorption, $CH_3$ chemical etching and eventually $H_{CH3}$ sputtering. The differences in energy ranges point back to the difficulty of describing quantitatively complex quantum phenomena with a classical potential. However, the agreement between classical and quantum approaches on the nature and sequence of mechanisms allows to further validate the classical C-H REBO potential with respect to our previous study [14]. It brings fundamental information about incident energy ranges and possible reaction mechanisms. Nevertheless, key differences have to be considered if one wants to investigate real $H_2$ plasma process conditions. First, the graphene sample would be randomly bombarded by both (i) $H/H_2$ neutral species impacting isotropically with thermal energies (300 K) and (ii) $H^+/H^{2+}/H^{3+}$ ions impacting anisotropically with higher energies (10–100 eV). Second, the graphene sample temperature is generally closer to 300K than 0K. Because thermal vibrations of the substrate carbon atoms may modify the reaction probabilities for reflection, etching or sputtering, we perform further CMD calculations of $H/CH_3$ interaction with higher surface temperatures in the next section.

## 4. Influence of surface temperature

We present in this part results from CMD simulations for cell temperatures of 300K and 600K since it is well known that plasma treatment usually takes place at room temperatures or higher. The goal of such a study is to show how the graphene and $CH_3$ residue temperature may modify the reaction probabilities for reflection, etching or sputtering. Due to thermal vibrations of both the graphene lattice and the $CH_3$ molecule, incident H species may not impact exactly on top of (or perpendicularly to) the $C_{CH3}$ atom of the target residue which is also a better description of the actual system.

The $CH_3$ radical absorbed on the graphene single layer is quenched along with the other substrate atoms to 300K and 600K (surface temperature $T_{cell}$) prior to H bombardment. Incident H energies $E_i$ are varied from 1 to 25 eV. For each ($E_i$, $T_{cell}$) combination, we perform 100 impacts at normal incidence on a refreshed $CH_3$ residue adsorbed on graphene (i.e. returned to its initial configuration). Then we calculate statistically the probability and energy thresholds for H reflection, $CH_4$ volatile product formation, $CH_3$ etching or $H_{CH3}$ etching from the target methyl radical.

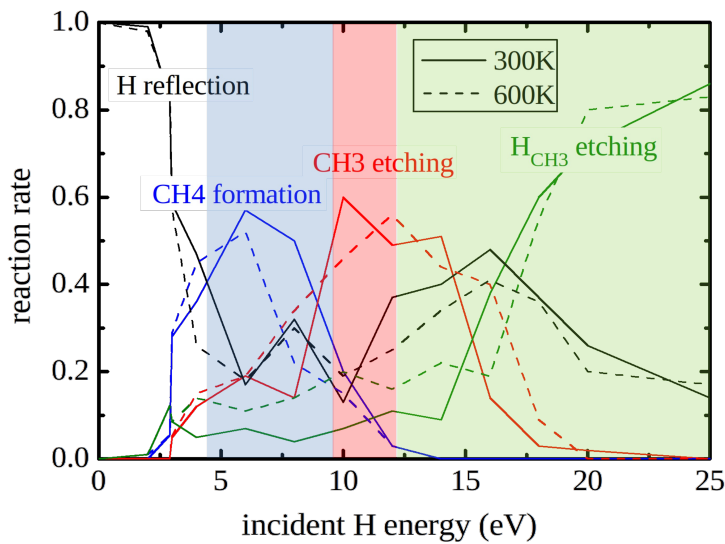

Figure 9 : Influence of H incident energy and graphene surface temperature on the probabilities of H reflection (black/white), $CH_4$ formation (blue), $CH_3$ etching (red) and $H_{CH3}$ etching (green) obtained from CMD statistical calculations. The colored zones indicate the exact energy ranges for these same mechanisms at 0K. The solid and dotted curves show the probabilities for each mechanism at 300K and 600K respectively.

Figure 9 shows the surface reaction probabilities predicted by CMD depending on the cell temperature and the incident H energy. The colored zones remind the energy ranges found for the different processes at 0K, as previously discussed. At 0K, H reflection takes place for $E_i$ < 3.7 eV due to the presence of a potential energy barrier of ~ 2.9 eV arising from the three H atoms belonging to the methyl radical. For $E_i$ ranging from 3.7 to 9 eV, $CH_4$ formation and desorption from the graphene surface occurs because H has enough energy to overcome the potential barrier and bind to the $CH_3$ radical (but not enough to induce the fragmentation of the molecule). For higher incident H energies in the [9-11.6] eV range, CMD shows that H can etch the $CH_3$ radical by forming a $CH_4$ volatile molecule which then splits into a single H atom and a $CH_3$ group, leaving the graphene surface clean and undamaged. Finally, for $E_i > 12$ eV, an excessive energy is transferred to the $CH_3$ residue during the impact, leading to the partial or full sputtering of $H_{CH3}$ atoms belonging to the $CH_3$ radical (leaving a stripped C atom chemisorbed on the surface).

As shown by the solid and dotted curves (corresponding to reaction rates for graphene surfaces at 300K and 600K respectively), rising the temperature of the graphene cell has a significant effect on all mechanisms described above. We observe that increasing the surface temperature broadens the energy ranges for which each mechanism - $CH_4$ formation, $CH_3$ etching or $H_{CH3}$ etching - can occur.

**4. Discussion**

Quantum and classical molecular dynamics studies at 0K predict the same sequence of reaction mechanisms. The thermal motion of the graphene structure at higher temperatures (300K and 600K) leads to stronger fluctuations of atomic positions, bond angles and bond lengths compared to the 0K case. As a consequence, for a same incident H energy, different reaction mechanisms can occur depending on the state/position of the $CH_3$ radical when the atomic H collides with it. Thus, an increase in the surface temperature lowers the potential barriers required for H chemisorption on the graphene surface [14] and less energy is needed to allow $CH_4$ desorption from a heated surface. Due to stronger vibrations of $C_{CH3}$- $H_{CH3}$ bonds at 300K and 600K, the energy range for which $CH_3$ etching can be observed is also wider. The latter also explains the occurrence of $H_{CH3}$ etching events along all the scanned energy range.

These calculations show that it is possible to etch $CH_3$ residues from graphene at room and higher temperatures with atomic hydrogen in the 3-15 eV energy range. In this range, H species can chemisorb on the graphene basal plane but have a low probability to damage the graphene structure (by breaking C-C bonds or creating vacancies) [14]. It may therefore be possible to clean graphene from simple $C_xH_y$ residues in low-temperature $H_2$ plasmas without making irreversible damages to the material.

**5. Conclusion**

The interaction of an incident energetic H atom with a $CH_3$ radical adsorbed on graphene was investigated by QMD and CMD calculations. We first simulated the elementary interaction between an incident atomic hydrogen and a $CH_3$ radical adsorbed on a graphene layer at 0K as a function of the incident H kinetic energy. Both QMD and CMD calculations showed the possibility for chemical etching of the methyl radical from graphene without damaging the graphene basal plane. In all cases, this happens through the formation of a $CH_4$ volatile compound that leaves the graphene surface either in the $CH_4$ form or breaking into a $CH_3$ group and an isolated H atom. While the energy ranges of occurrence of the mechanisms differ between QMD and CMD, both approaches give the same sequence for increasing energies: incident H reflection, $CH_4$ formation with $CH_4$ or $CH_3$ + H leaving the graphene basal plane. Results only slightly differ at high energy where both approaches show a reflection of the incident H atom, but with an eventual sputtering of the $H_{CH3}$ atoms belonging to the $CH_3$ residue in the case of CMD which is not found in QMD. This last difference and the variation in the energy ranges associated to each mechanism point back to the difficulty to describe so complex phenomena and especially both $sp^2$ and $sp^3$ hybridization in carbon. Nevertheless the agreement on the sequence of mechanisms further validates the REBO potential.

Temperature and incident H energy influence on surface reactions probabilities were discussed in the last part of the article. It was shown that increasing the surface temperature broadens the energy ranges for which each mechanism can occur. Indeed the potential barriers for H adsorption are lowered and less energy is needed to allow $CH_4$ desorption from a heated surface.

A $CH_3$ radical absorbed on graphene is a "basic" residue representation but these results are interesting for the development of future graphene surface cleaning processes by $H_2$ plasmas. Indeed, there is very little information regarding graphene cleaning by plasmas in the literature and as a general rule, the incident kinetic energy of radicals and ions required to selectively etch residues is unknown. Similarly, PMMA etching by $H_2$ plasmas has not been deeply investigated. However, it was shown that downstream plasmas (in which the surface is bombarded only by low energy H radicals) can be used to clean the PMMA without damaging the graphene at high temperature (H adsorption is reversible by thermal annealing). According to our work, the contaminant carbon was probably removed as volatile $CH_4$ under these conditions. However, we underline that, from our results, $CH_3$ residues can be removed from the surface by low-energy ions in the 1-15 eV range without damaging the graphene irreversibly. This suggests that high density plasma sources (such as Inductively-Coupled and Electron Cyclotron Resonance plasmas) could also be advantageously used for this purpose: by contrast with downstream systems, they provide large fluxes of reactive radicals and ions and are thus expected to be much more efficient to clean graphene.

## 6. Acknowledgement

The authors wish to thank the Nanosciences Foundation of Grenoble in the frame of the 2010 Chairs of Excellence Program and the NANOSIM-GRAPHENE project ANR-09-016-01 for financial support. This work was partly supported by the french RENATECH network. QMD simulations have been performed at the GENCI computational facility (Grant 097015).